\documentstyle[pre,multicol,aps,graphicx,psfig]{revtex}

\begin{document}

\title{Finite-size effects in the self-organized critical forest-fire model}

\author{Klaus Schenk$^1$, Barbara Drossel$^2$, Siegfried Clar$^3$, Franz
Schwabl$^1$
}
\address{${}^1$ Physik-Department der Technischen Universit\"at M\"unchen, James
Franck Stra\ss e, D-85747 Garching}
\address{${}^2$ Theory Group, Department of Physics and Astronomy, University of
Manchester, Manchester M13 9PL, UK}
\address{${}^3$ iXOS Software AG, D-85630 Grasbrunn, Germany }
\date{\today}
\maketitle

\begin{abstract}
  We study finite-size effects in the self-organized critical
  forest-fire model by numerically evaluating the tree density and the
  fire size distribution. The results show that this model does not
  display the finite-size scaling seen in conventional critical
  systems.  Rather, the system is composed of relatively homogeneous
  patches of different tree densities, leading to two qualitatively
  different types of fires: those that span an entire patch and those
  that don't. As the system size becomes smaller, the system contains
  less patches, and finally becomes homogeneous, with
  large density fluctuations in time.
\end{abstract}

\begin{multicols}{2}
\narrowtext
\section{Introduction}
During the past years, systems which exhibit self--organized
criticality (SOC) have attracted much attention, since they might
explain part of the abundance of fractal structures in nature
\cite{bak87}. Their common features are slow driving or energy input
and rare dissipation events which are instantaneous on the time scale
of driving.  In the stationary state, the size distribution of
dissipation events obeys a power law, irrespective of initial
conditions and without the need to fine-tune parameters. Examples for
such systems are the sandpile model \cite{bak87}, the self-organized
critical forest fire model \cite{dro92,cla94}, the earthquake model by
Olami, Feder, and Christensen \cite{ola92}, and the Bak-Sneppen
evolution model \cite{sne93}. Numerical as well as analytical studies
of those systems are usually based on the assumption that their
critical behaviour can be described in similar terms as that of
equilibrium critical systems.  This assumption is given a basis in
\cite{sor95}, where it is suggested that SOC systems can be mapped on
conventional critical systems by interchanging control and order
parameters. Thus, the Bak-Sneppen model can be mapped on a depinning
problem \cite{sne93}.  However, it has been shown in \cite{cla97} that
the mapping suggested in \cite{sor95} for the SOC forest-fire model
does not generate a system with a conventional critical
point. Instead, the phase transition shows hysteresis effects and is
discontinuous when approached from above. Other unconventional
features have also been seen in the SOC forest-fire model, like the
existence of more than one diverging length scale \cite{cla95,hon96},
the absence of a spanning cluster immediately beyond the critical
point \cite{cla95}, and the dependence of the large-scale behaviour on
details of the model rules \cite{hon96,dro96}. (For a review on the
SOC forest-fire model, see \cite{cla96}.) Other SOC systems show also
unconventional scaling behaviour. Thus, in the two-dimensional abelian
sandpile model finite-size scaling is violated \cite{teb99}, and the
critical exponents for the earthquake model by Olami, Feder, and
Christensen \cite{ola92} appear to depend continuously on the
parameters. There is substantial need to better understand the nature of
the scaling behaviour of those systems.

 It is the purpose of this paper to shed some light on the
unconventional critical behaviour of the SOC forest-fire model by
studying its finite-size effects. We choose a version of the model
which is identical to the SOC forest-fire model for system sizes much
larger than the correlation length, and we discuss the changes that
occur in the model as the system size is decreased below the
correlation length. We find that instead of displaying finite-size
scaling, small systems undergo a rearrangement from a structure with
patches of different density to a more homogeneous structure with
large density fluctuations in time. We find also that, contrary to
conventional critical systems, small systems and small parts of large
systems differ in the probability distribution for the density and in
the fire-size distribution.  We suggest that these results can be
explained by the fact that the system has two qualitatively different
types of fires.

The outline of this paper is as follows: In section II, we define the
model that we used for studying finite-size effects, and discuss
briefly known results. Section III shows computer simulation results
for the fire-size distribution and the tree density as the system size
changes from values larger than the correlation length to values much
smaller than it. In the conclusion, we summarize and discuss our
findings.

\section{The Model}
\label{model}
The version of the SOC forest-fire model studied in this paper is
defined on a square lattice with $L^2$ sites. Each site is either
occupied (``tree'') or empty (``no tree'').  At each time step, the
system is updated according to the following rules: (i) ``Burning'': A
site in the system is chosen at random (``struck by lightning''). If
the site is occupied, the whole cluster of occupied sites connected to
this site (by nearest-neighbour coupling) is removed from the system
(``burnt''), i.e., the occupied sites of that cluster turn to empty
sites. If the chosen site is empty, nothing happens. (ii) ``Tree
growth'': We select randomly $s_0\equiv pL^2$ sites from the system
and occupy those that are empty (possibly also including sites which
have become empty due to the removal of the cluster). These sites are
selected one after another, allowing for the same site being selected
more than once during the same filling step. In principle, $s_0$ can
therefore be larger than $L^2$, however, in our simulations we chose
usually values smaller than $L^2$.

For fixed $s_0$ and very large system size $L$, these rules are
equivalent to having a lightning probability $f=1/L^2$ per site and
time step, and a tree growth probability $p$, and the model is
identical to the original SOC forest-fire model \cite{dro92}. Because
of this equivalence, which was first pointed out by Grassberger
\cite{gra93} most numerical studies of the SOC forest-fire model up to
now were performed using the above rules, which allow for fast and
efficient computer simulations. With the above rules, finite-size
effects can also be studied very efficiently, as was suggested in
\cite{hon96}. However, one has to keep in mind that the results are
somewhat different from those for the original model. While in the
original model lightning can strike the system between the growth of
any two trees, it can strike the system in the present model only
after growth step (ii) is finished. This leads to density peaks in
Figures~\ref{bild9} and \ref{bild10} below that are not present in the
original model. However, our main conclusions are not affected by the
particular choice of the dynamical rules, as will be discussed further
below.

Let us first summarize shortly the major numerical results for the case
$s_0 \ll L^2$, as reported in the literature on the SOC forest-fire
model\cite{cla94,gra93,hen93,chr93}. In this limit, only a small
number of trees grow at each time step (compared to the total number
of trees). After a transient time, a stationary state is reached where
the tree density has only small fluctuations around some average value
$\bar \rho(s_0)$ that does not depend on $L$. Throughout this paper,
we study only stationary states and do not evaluate the initial
transient behaviour. Since the mean number of trees $\bar s$ burnt
during a fire must be identical to the mean number of trees growing
between two fires, we have the relation
\begin{equation}
\bar s = s_0(1-\bar \rho)/\bar \rho.
\end{equation}
The leading finite-size corrections to this equation are of order
$s_0/L^2$ and can be neglected in the case $s_0 \ll L^2$ which we are
considering in this paragraph.  As $s_0$ increases, the mean fire size
increases also, and we approach the critical point of the SOC
forest-fire model, where the mean tree density is given by $\bar
\rho_c \simeq 0.41$. The correlation length $\xi$ is a measure for the
radius of the largest tree cluster and is related to $s_0$ via $\xi
\sim s_0^{\nu}$, with $\nu \approx$ 0.58 in $d=2$ dimensions.  The
size distribution of tree clusters near the critical point is well
described by the scaling form
\begin{equation}
n(s) \simeq s^{-\tau}{\cal C}(s/s_{max})\, , \label{scaling}
\end{equation}
with a cutoff function ${\cal C}$ that is constant for small arguments
and decays exponentially fast when the argument is considerably larger
than 1. The cutoff cluster size $s_{max}$ is related to the
correlation length $\xi$ via $s_{max} \sim \xi^\mu$, with $\mu$ being
the fractal dimension of tree clusters, which is found to be 1.95 \cite{hen93}
or 1.96 \cite{cla94,sch99}. The value of the exponent $\tau$ is
approximately 2.14. The relation between $s_{max}$ and $s_0$ is
$s_{max} \sim s_0^\lambda$, with $\lambda = \nu\mu \simeq 1.15$
\cite{cla94}.

All these numerical findings agree well with conventional scaling
assumptions based on a single diverging length scale.  Analytical
studies of the model, such as mean-field theories
\cite{chr93,dro94c,ves98} and renormalization group calculations
\cite{pat94,lor95} are also based on conventional scaling assumptions.
Therefore, the violation of finite-size scaling described in the
following might appear surprising to many readers.  However, one must
keep in mind that the simulation data do not cover much more than one
decade in the correlation length $\xi$. The observed scaling behaviour
eq.~(\ref{scaling}), together with the measured values of the critical
exponents, do not necessarily indicate an exact asymptotic scaling
form, but may simply be a good approximation to more complicated
scaling, which works well for the system sizes and parameter values
studied in simulations. A similar phenomenon is known for the sandpile
model, where good scaling collapses for the avalanche size
distribution could be achieved in \cite{lub97,chessa} and older
papers, although it has been recently shown \cite{teb99} that
finite-size scaling is violated and that the simple scaling ansatz
used for the data collapse is incorrect.

\begin{figure}
\psfig{file=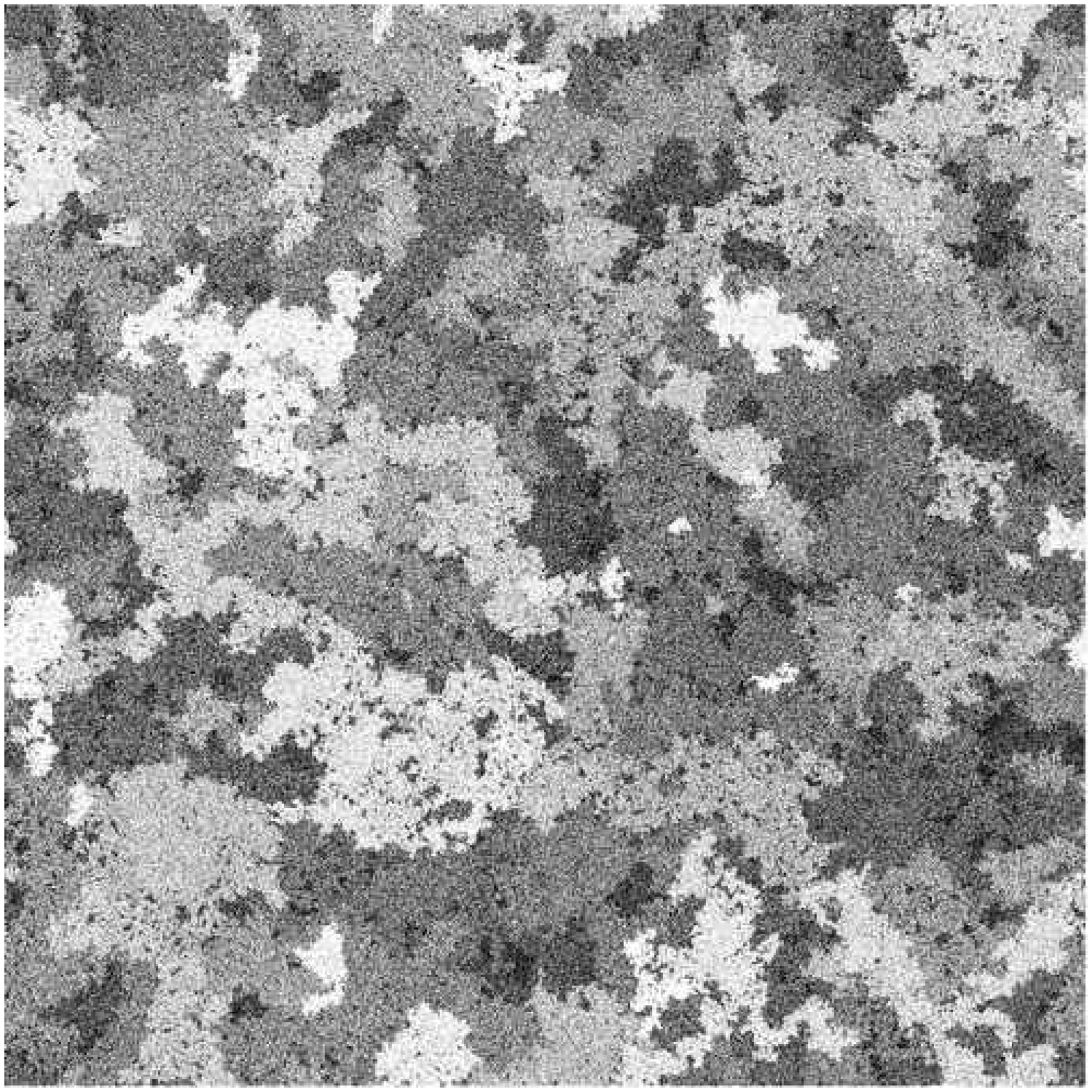,width=0.46\textwidth,angle=0}
\caption{Snapshot of the SOC forest-fire model for $\bar{\rho} \simeq
\bar{\rho_c} \simeq$ 40.8\% and $L$=4096. Trees are black and empty sites
are white.} \label{bild1}
\end{figure}

Figure~\ref{bild1} shows a snapshot of a system with a tree density
$\bar \rho$ just below $\bar \rho_c$. One can distinguish regions of
different densities with a rather homogeneous tree distribution within
a region. These regions are obviously created by a fire that burns
down a cluster of high tree density. After the fire, a burnt region is
almost empty and becomes slowly filled with trees according to the law
$\dot \rho = p(1-\rho)$. We call these regions of homogeneous tree
density ``patches'', as we did in \cite{cla97a}.  If lightning strikes
a tree in a patch of low density, it usually burns down a small tree
cluster. If it strikes a patch of a density larger than the
percolation threshold, it burns down a tree cluster as large as the
patch itself. This observation indicates that there are two
qualitatively different types of fires in the system: those that span
an entire patch, and those that destroy a small percolation cluster
within a patch of a tree density below the percolation threshold. As
we will see below, this gives rise to the unusual finite-size
properties of the model.

If the correlation length $\xi$ is of the same order as or larger than
$L$, the behaviour sketched above is modified due to finite-size
effects. For not too small values of $s_0/L^2$, the tree density
increases by a noticeable amount between two fires, leading to large
density fluctuations and to fires that span the entire system. If the
SOC forest-fire model showed conventional critical behaviour, there
would be a single diverging length scale, namely the correlation
length $\xi$, which would be related to $f/p$ or, equivalently, to
$s_0$, via $\xi \sim (f/p)^{-\nu}$ or $\xi \sim s_0^\nu$. Finite-size
effects would then manifest themselves in a scaling form
\begin{equation}
n(s) \simeq s^{-\tau}{\cal C}(s/L^\mu)\, , \label{FSS}
\end{equation}
for the size distribution of tree clusters. Furthermore, on scales
smaller than $L$ and $\xi$, all measured quantities should be
indistinguishable from those measured in a small section of an
infinitely large critical system.

The following section presents simulation results that show that none
of these finite-size scaling assumptions is satisfied for the SOC
forest-fire model. In fact, the invalidity of the assumption of a
single diverging length scale has already been shown in
\cite{hon96}. The invalidity of the second assumption that
measurements in small systems and in small sections of large systems
should give identical results, can be understood by considering for
instance the mean time interval between two fires. In a small
subsystem of linear size $l$ of a large system with a correlation
length $\xi \gg l$, this is given by $(p(1-\bar
\rho_c))^{-1}$. Just before fire reaches the subsystem, its tree density is
far above the percolation threshold, and the spanning cluster of the
subsystem is part of a large tree cluster that extends far beyond the
limits of the subsystem. Lightning usually strikes this large cluster
outside the subsystem, the time interval between two lightning strokes
within the subsystem being $L^2/ l^2$, which diverges as $L$
diverges. In contrast, fire cannot enter a small system from outside,
but the tree density of a small system increases until lightning
strikes a tree within the system. According to our rules, time is
measured in units of the mean time interval between two lightning
strokes. On this time scale, the time between two fires within a small
system of linear size $l$ is finite. In contrast, the time interval
between two fires within a subsystem of size $l$ of a much larger
system is vanishingly small compared to the time interval between two
lightning strokes within the subsystem. All these arguments are backed
up and complemented by the numerical results reported in the
following.

\section{Results of Computer Simulations}
In this section we will present and explain data obtained from about
300 runs of the model for various values of $s_0$ and $L$. Since
many runs of the simulation were necessary, we chose a
cluster of workstations rather than a "supercomputer". The system size
$L$ varied between 10 and 2000 in these runs. We found that as
finite-size effects become more important, the system shows a
transition between two qualitatively different types of behaviour
which we call critical behaviour and percolation-like behaviour. The
critical behaviour is is characterized by a good scaling collapse of
the fire size distribution and by large spatial variations in the
local tree density. The percolation-like behaviour is characterized by
large temporal fluctuations in the global tree density, with a rather
homogeneous tree distribution within the system for any given
time. Snapshots of the system therefore resemble percolation systems
where each site is occupied by a tree with a probability $\rho$. (For an introduction to percolation theory, see e.g. \cite{sta92}.) The
following three subsections show how this transition manifests itself
in the mean tree density, the fire size distribution, and the
probability distribution for the tree density.

\subsection{Lines of constant tree density}

First, we measured the mean tree density
\begin{equation}
\bar \rho = (1/T) \sum_{t=1}^T \rho(t)\label{rho}
\end{equation} 
in the system, averaged
over a large number of $T$ iterations, for various values of $L$
and $s_0$. The density $\rho(t)$ was always evaluated after the
refilling step (ii). Compared to a model where trees grow at a
rate $\dot \rho = p(1-\rho)$, the density values in our
model are somewhat larger when $s_0/L^2$ is not very small. If, for
instance, the density is increased from $\rho - \Delta \rho$ to $\rho$
during the refilling step (ii), the value $\rho$ enters the above
sum, while an evaluation based on a constant growth rate would give
$1-\Delta \rho/\ln(1+\Delta\rho/(1-\rho))$ instead of $\rho$.

It it not obvious what the relation between $s_0$, $L^2$ and $\bar
\rho$ should be if we want to deduce it from an analogy with
equilibrium critical systems. We have already mentioned that the
temporal fluctuations in $\rho(t)$ become larger as the ratio
$s_0/L^2$ increases. Similarly, temporal fluctuations increase in a
critical equilibrium system when the system size becomes smaller. One
might therefore expect that decreasing $L$ at fixed $s_0$ should drive
the system toward the critical point, where $\bar \rho = \bar \rho_c
\simeq 0.41$.  However, we have argued in the previous section that a
given site burns down more often in a large system without finite-size
effects than in a smaller system with finite-size effects that has the
same value of $s_0$. From this, it follows that the mean tree density
increases with decreasing $L$, when $s_0$ is fixed. In the limit $s_0
\gg L^2$ it must go to one. From this point of view, a system with
sufficiently large finite-size effects should rather be compared to an
equilibrium system in the ordered phase, for instance to a percolation
system beyond the percolation threshold. The analogue of $s_0$ in a
percolation system is then the mean size of the cluster that a given
site belongs to, and it is proportional to $L^2$ beyond the
percolation threshold \cite{sta92}. Indeed, we find that the mean size
of fires becomes proportional to $L^2$ when finite-size effects are
strong (see below). However, there is nevertheless a fundamental
difference between a percolation system beyond the percolation
threshold and our system with a density above $\rho_c$: In our model,
a tree cluster that spans the system and has a size proportional to
$L^2$ occurs only rarely when the mean density is only slightly above
the critical density, while a percolation system above the percolation
threshold has always a system spanning tree cluster.

\begin{figure}
\includegraphics[width=0.46\textwidth,angle=0]{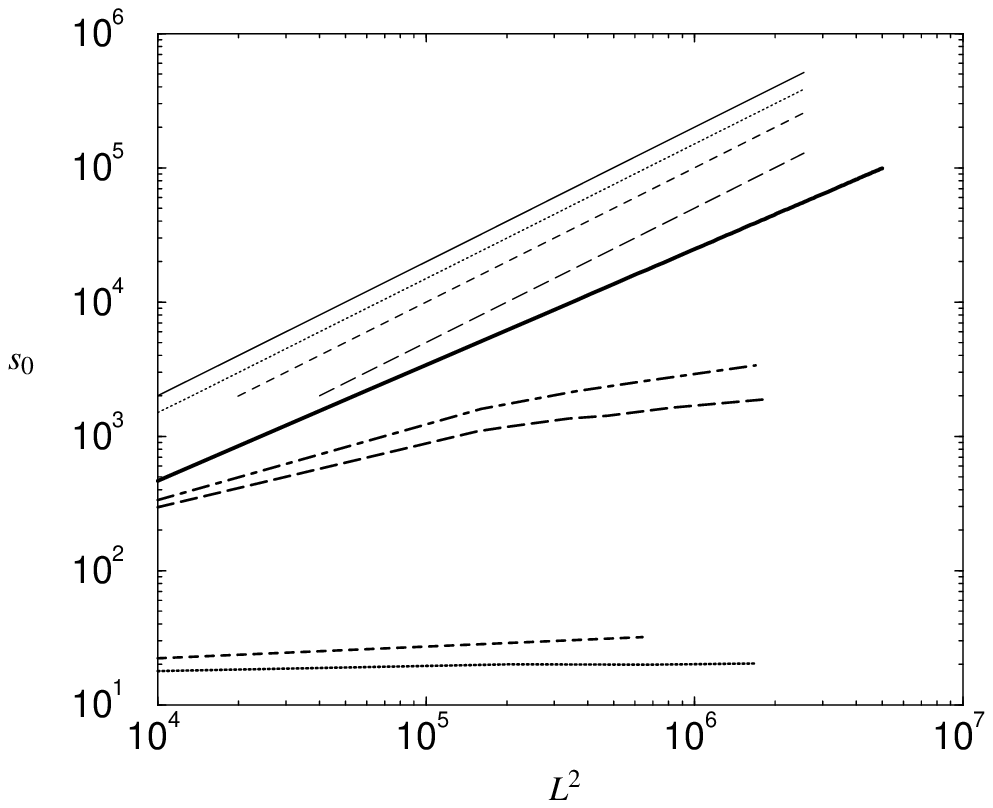}
\caption{Lines of constant mean tree density in a  $s_0$ vs. $L^2$ plane.
The bold solid line represents the separatrix between the SOC and
the percolation-like behaviour. The remaining lines represent
constant $\bar{\rho}=0.47,0.455,0.43,0.42,0.0403,0.40,0.35,0.343$ (from
top to bottom). The line for $\bar{\rho}=0.40$ is derived from
interpolated results. } \label{bild2}
\end{figure}

In Figure~\ref{bild2}, lines of constant mean tree density are plotted
on a double logarithmic scale in $s_0$ and $L^2$. This figure shows
that there are two qualitatively different regions in the $s_0$ vs. $L^2$
plane, 
with a transition region between them.  First, there is the region
where there are no finite-size effects, $s_0 \ll L^2$. In this region,
the size of the tree clusters is much smaller than the system size,
and the global density fluctuations are small. The mean tree density
of such a system is smaller than $\bar \rho_c \approx 0.41$. For
systems with small density fluctuations the probability that a given
empty site is filled with a tree during one time step, is given by
$s_0/L^2$, and the probability that a given tree is burnt by a fire is
$s_0 (1-\bar \rho)/(\bar \rho L^2)$. Since both probabilities decrease
as $1/L^2$ with increasing $L$, dynamics of larger systems are slower
than those of smaller systems. Apart from this change of the
characteristic time scales, the local dynamics is independent of $L$,
and consequently correlation functions and cluster size distributions
are the same for systems of different sizes (provided that $L^2 \gg
s_0$). This leads to the horizontal slope in the large $L$ regime of
the curves for $\bar \rho<\bar \rho_c$ in Figure~\ref{bild2}.  

The curves show deviations from the horizontal behaviour when $L$
becomes as small as or smaller than the correlation length $\xi \sim
s_0^\nu$.  These deviation occur in the transition region where
finite-size effects begin to become noticeable. The dynamics changes
from fires that are not affected by the finite system size to a fire
size distribution that includes sometimes events of the order of the
system size.  Fires of such a large size destroy the patchy structure
of the forest described earlier, and cause a more random tree
distribution.

In the second region, lines of constant mean density are curves of
constant $s_0/L^2$. This feature can best be understood when
considering the parameter range where $s_0$ is of the order of $ L^2$
or above. In this range, the mean tree density is much larger than
$\bar \rho_c$, and the system contains a spanning cluster after each
filling. During the "burning" step, this cluster is removed with a
finite probability, leading to a large change in density in the
system. When a "finite" (i.e. not spanning) cluster is removed during
the "burning" step, the overall density hardly changes for large
$L$. The time series of the density is therefore determined almost
completely by the filling events and by the large burning
events. Since the filling events fill a large fraction of empty sites,
and since large burning events burn a large fraction of trees, the
tree distribution within the system is rather homogeneous. A snapshot
of the system at a given time looks therefore similar to a percolation
system with a density $\rho(t)$. From percolation theory we know that
for a given density the fraction of trees sitting in the spanning
cluster is independent of $L$, and consequently curves of constant
large density are curves of constant $s_0/L^2$ in Figure~\ref{bild2}.
Even curves for smaller $s_0/L^2$, which correspond to densities only
slightly above $\bar \rho_c$ show for sufficiently large $L$ an
asymptotic behaviour $s_0/L^2 =$ const, with the constant vanishing at
$\bar \rho_c$. The reason is again that finite fires do not reduce the
density of an infinitely large system, and that the system spanning
fires reduce the density by an amount that does not depend on $L$, but
only on the density itself. 

Finally, the critical curve (for the density $\bar \rho= \bar \rho_c
\simeq 0.41$) is obtained from the condition
\begin{displaymath}
L \approx \xi \sim s_0^{\nu}, \; \; \mbox{with} \; \; \nu\simeq 0.58 \;.
\end{displaymath}
This curve is the separatrix between the two regions
described above and is indicated in Figure~\ref{bild2} by the bold
line.
As $L$ decreases, more and more curves merge
with the separatrix when their correlation length
becomes comparable to the system size.

\subsection{The fire size distribution}

Since lightning strikes each tree with the same probability, the size
distribution of fires is proportional to $sn(s)$, with $n(s)$ being
the size distribution of tree clusters. As mentioned above,
conventional scaling would imply a form $s n(s) \simeq s^{1-\tau}{\cal
C}(s/s_0^{\mu\nu})$ for the fire size distribution if the correlation
length $\xi \sim s_0^\nu$ is smaller than the system size, and a
finite-size scaling form $s n(s) \simeq s^{1-\tau}{\cal C}(s/L^\mu)$
in the opposite case. In both cases, one would obtain a scaling collapse of
the curves for different $s_0$ or $L$.

\begin{figure}
\includegraphics[width=0.46\textwidth,angle=0]{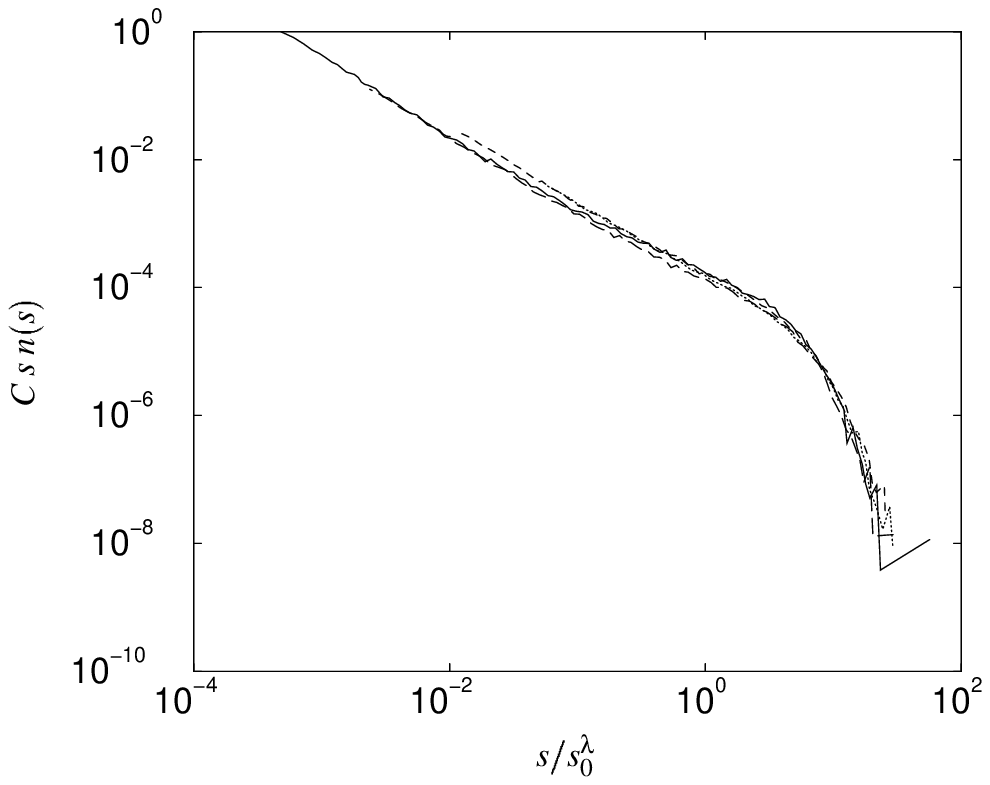}
\caption{Sca\-ling col\-lapse of the fire size dis\-tri\-bu\-tion
for sys\-tems with the parameters ${s_0}/{L^2}=0.001$ and 
$L=$ $1600$, $800$, $400$, $200$. The measured mean tree densities 
are $\bar{\rho}=$ $0.40$, $0.395$,
$0.38$, $0.36$. $C$ is a suitable scaling constant for each curve.} \label{bild4}
\end{figure}

Figure ~\ref{bild4} shows the fire size distribution for parameters
such that $\xi < L$. While not perfect, the data collapse is good and
would not impose the conclusion that simple scaling is violated.  The
bump near the end of the curves indicates that the cutoff function
${\cal C}$ increases first with increasing argument, before it shows
the exponential decay. This bump is believed to contain all the trees
that would sit in larger clusters if the system was exactly at the
critical point \cite{gra93}.

\begin{figure}
\includegraphics[width=0.46\textwidth,angle=0]{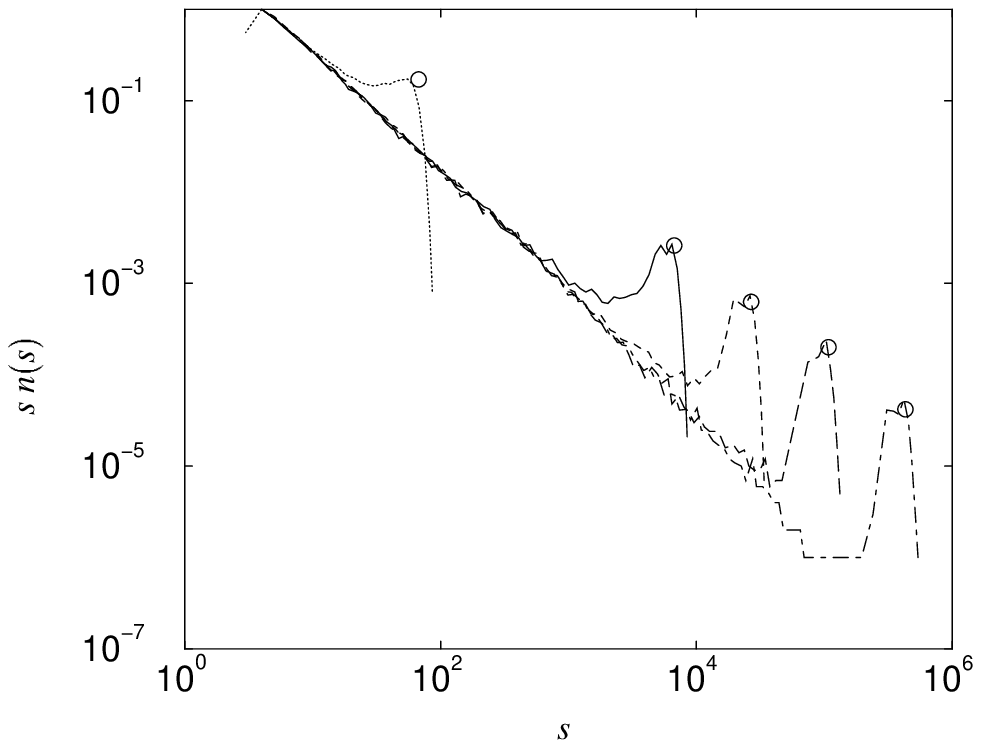}
\caption{The size
distribution of fires for ${s_0}/{L^2}$=0.15 and $\bar
\rho=0.454$, for the system sizes $L=10, 100,200,400,800$ (as the
peaks move from left to right). The circles mark the points of
exact $L^2$ scaling, taking the peak of the $L=800$ curve as
reference.} \label{bild5}
\end{figure}

Figure ~\ref{bild5} shows the fire size distribution for parameter
values such that the mean density is ten percent above its critical
value. As discussed in the previous subsection, system spanning fires
occur, and their size scales as $L^2$.  These fires are responsible
for the peaks in the fire size distribution. Similar peaks occur in
equilibrium critical systems in the ordered phase, for example in the
cluster size distribution of a percolation system above the
percolation threshold. In a percolation system, the occurrence of such
peaks implies that the system size is larger than the correlation
length, which is identical to the cutoff in the radius of the finite
(i.e., non system spanning) clusters. In our system, however, we do
not see such an exponential cutoff to the size distribution of the
finite clusters. Instead, the curve for $L=800$ in figure ~\ref{bild5}
appears to obey a power law from $s \simeq 100$ up to the point where
the peak begins. The explanation for this unusual behaviour must lie
in the large temporal fluctuations in the density. The density is only
sometimes so large that the large fires, which have a size of the
order $L^2/2$, occur. At other times, the density values are different
and allow for a broad range of other fire sizes.

\begin{figure}
\includegraphics[width=0.46\textwidth,angle=0]{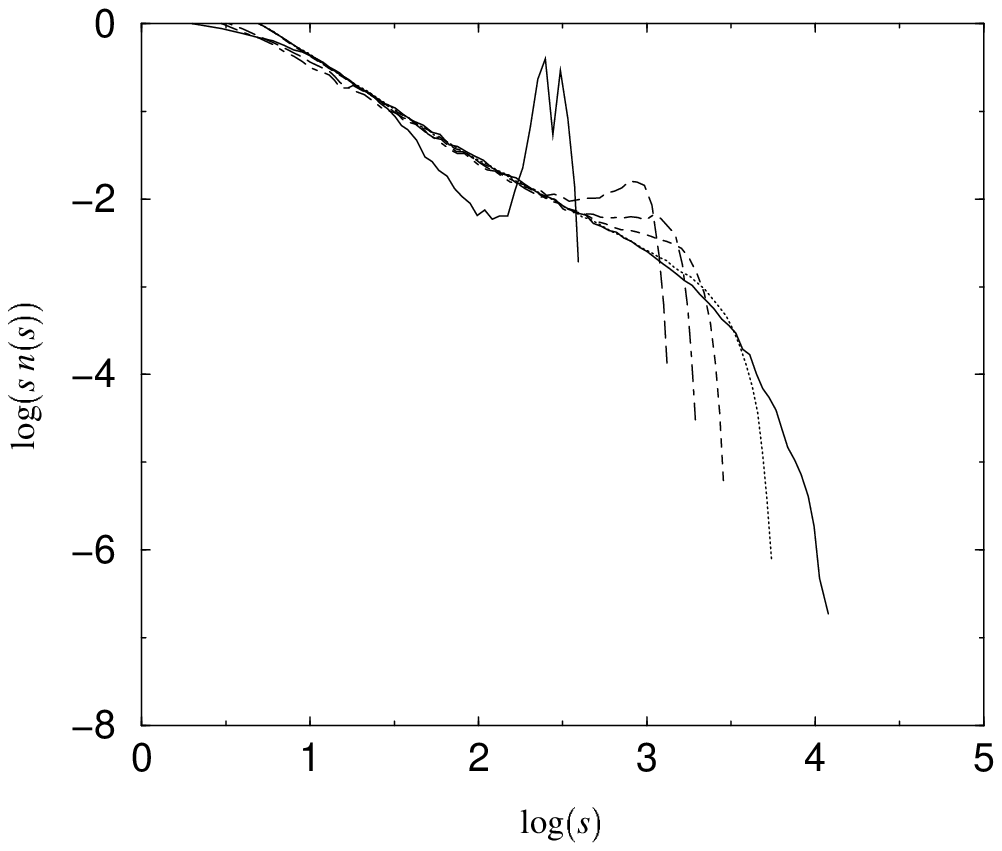}
\caption{Transition from critical to percolation-like behaviour as the
system size $L$ is decreased for the fixed parameter $s_0$=200. The
parameter $L$ of the curves are from right to left: $L=$ $1300$,
$100$, $63$, $50$, $40$, $20$. The measured mean tree densities are
$\bar{\rho}=$ $0.385$, $0.392$, $0.402$, $0.414$, $0.432$, $0.577$.}
\label{bild6}
\end{figure}

The transition between critical scaling and $L^2$-scaling can be
observed when $L$ is varied for fixed $s_0$, as illustrated in
Figure~\ref{bild6}. One can see that the shape of the curves changes
continuously as $L$ is decreased. Clearly, because of this change in
shape, finite-size effects do not manifest themselves in a scaling
behaviour $s n(s) \simeq s^{1-\tau}{\cal C}(s/L^\mu)$. It is
impossible to generate a scaling collapse of different curves, even if
their density is close to the critical density. Furthermore, as
mentioned above, the cutoff introduced by the finite system size
always scales as $L^2$, due to the occurrence of system spanning
fires, and not as $L^\mu$, as expected for conventional critical
systems. 

For small
system sizes, spanning clusters may already occur for densities below
$\bar \rho_c$, an effect which is clearly visible in the curve for
$L=63$. The formation of peaks due to finite
size effects, was also found in~\cite{mala98}.

\begin{figure}
\includegraphics[width=0.46\textwidth,angle=0]{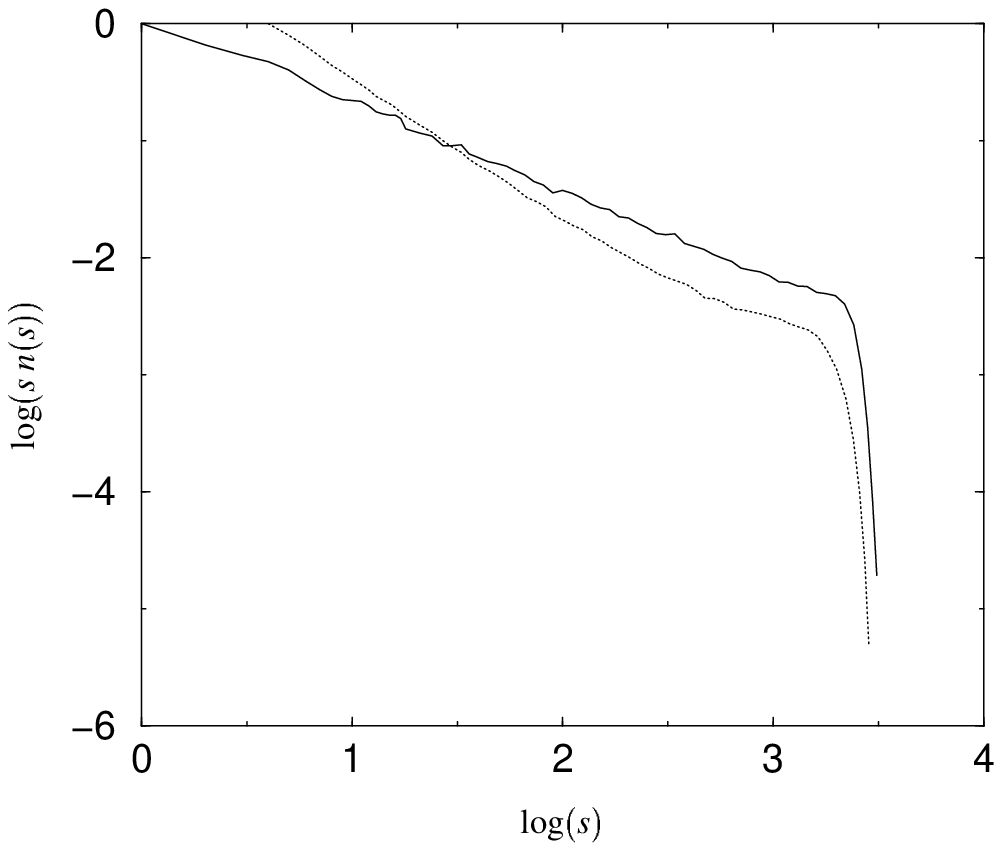}
\caption{Comparison of the fire size distribution of a section of
size $l=$63 of a larger ($L=$600, $s_0=$1440) system (solid line) and a small system with
$L=$63 ($s_0=$200) (dotted line). For both  systems we measured a mean  tree density of
$\bar{\rho}=$0.401.} \label{bild6b}
\end{figure}

As mentioned earlier, for conventional critical systems a
system of small size and a small section of a large system are
equivalent.  We have argued that this is not true for the forest-fire
model since the mean tree density and the time interval between fires
are different in the two cases.  The next two figures show that also
the fire size distributions are different.  In Figure~\ref{bild6b} the
fire size distribution of a section of a large system and a
corresponding small system is shown. The scaling parts and the form of
the bumps near the cutoff are very different.  The fire size
distribution of a small section of a large system is broader than that
of a small system. The reason is that a section of a large system can
contain a boundary between a patch of large tree density and a patch
of small tree density.  This boundary can pass through the section in
different ways, and the number of trees in the dense part can take
different values.  Since large fires only burn the dense part, the
size distribution of fires becomes broad.

\begin{figure}
\includegraphics[width=0.46\textwidth,angle=0]{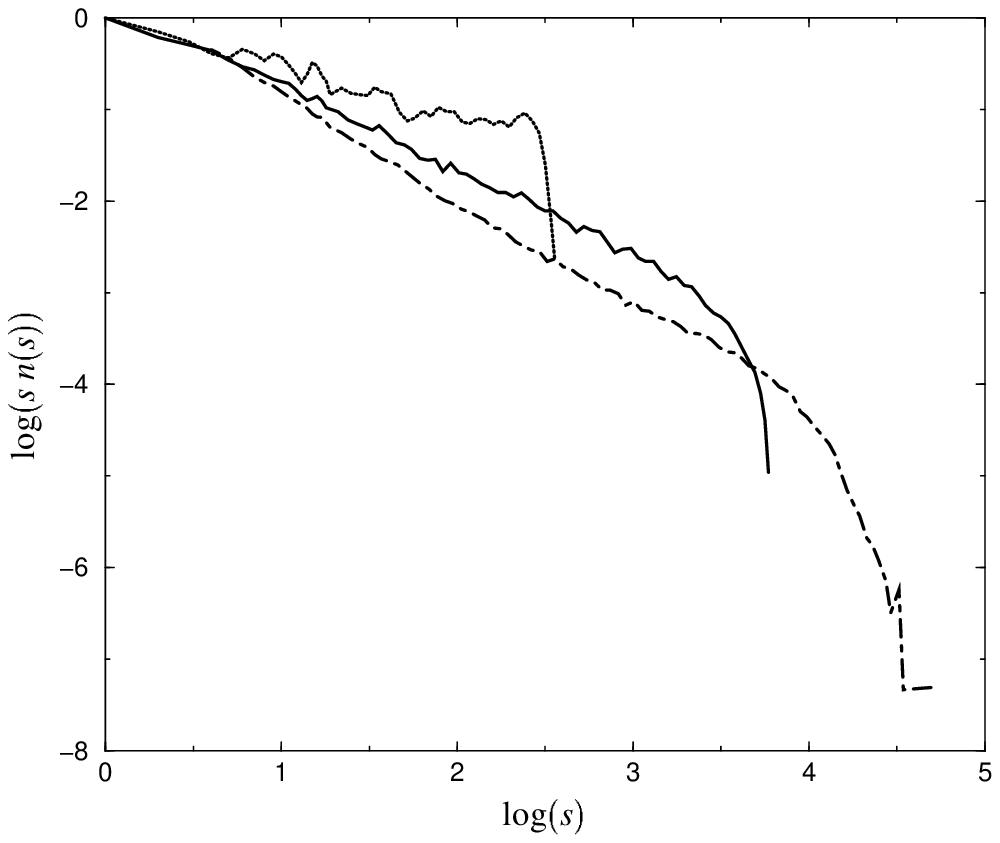}
\caption{Fire size distribution for systems with $L=$ 800 (dot-dashed
line) and of subsections of this system of size $l=$100
(solid line) and $l=$20. We used ${s_0}/{L^2}=0.001$ and measured the  mean tree density 
 $\bar \rho \approx 0.40$.} \label{bild6a}
\end{figure}

Figure~\ref{bild6a} shows how the fire size distribution changes when
smaller and smaller sections of a large system are evaluated.
Comparing to Figure~\ref{bild6}, one sees again that the fire size
distribution of small sections of large systems is different from that
of small systems.

To summarize this subsection, the fire size distribution in the
presence of finite-size effects does not show the features of
finite-size effects in conventional critical systems. We find a
continuous change in the shape of the fire size distribution and
cutoffs that scale as $L^2$, rather than conventional finite-size
scaling. Furthermore, the fire size distribution in small sections of
large systems is different from small systems.  Our results for the
fire-size distribution confirm the qualitative transition from a
parameter region unaffected by finite-size effects to a region
dominated by system spanning fires that we found in the previous
subsection.

\subsection{Probability distribution of the density}

Finally, we studied the temporal fluctuations in the values of tree
density $\rho(t)$ by measuring how often a given value of $\rho$
occurs within a sufficiently long time series.  We denote by
$w(\rho)d\rho$ the probability that the tree density lies in the
interval between $\rho$ and $\rho+d\rho$. The quantity $w(\rho)$ is
therefore the probability density for the tree density $\rho$.  We
measured $\rho$ always after the trees were refilled, i.e., after step
(ii). The results show again a qualitative change as the correlation
length becomes smaller than the system size, reflecting the transition
from critical to percolation-like behaviour.

\begin{figure}
\includegraphics[width=0.46\textwidth,angle=0]{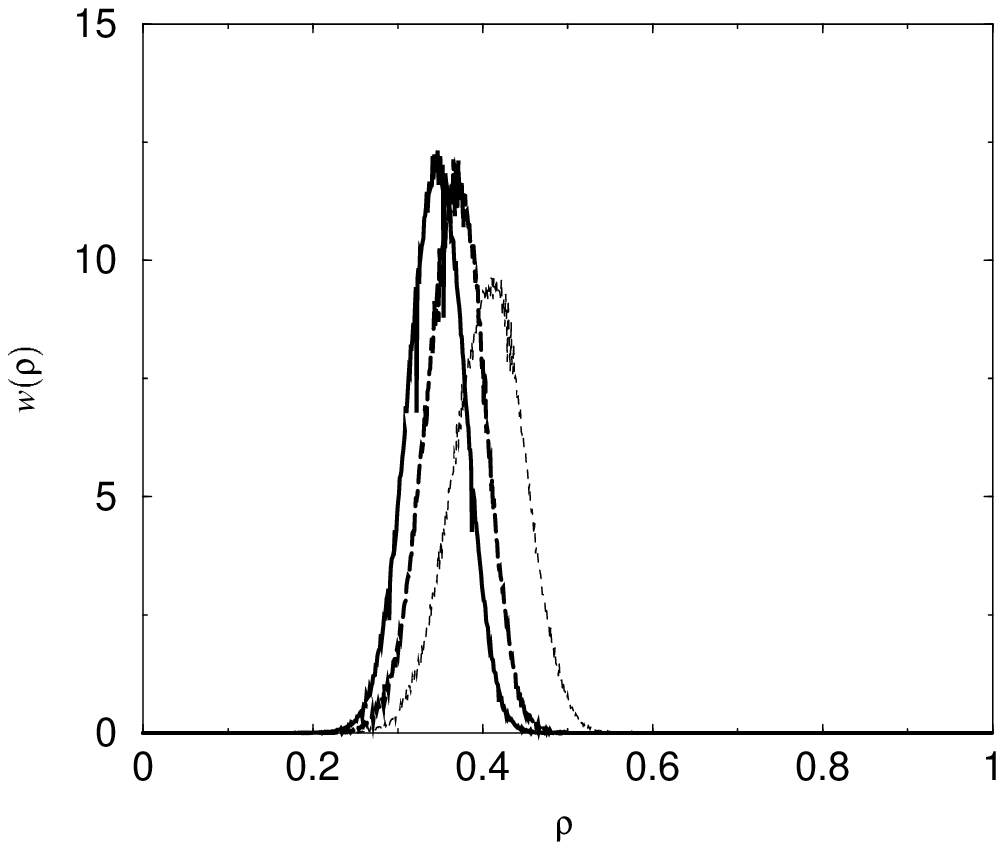}
\caption{$w(\rho)$ for sys\-tems with
the parameters ${s_0}/{L^2}$=0.005 and  $L=$ $63$, $100$, $1600$
(as the peaks move from left to right). The measured mean tree  densities 
are $\bar{\rho}=$
$0.345$, $0.367$, $0.407$.
} \label{bild7}
\end{figure}

\begin{figure}
\includegraphics[width=0.46\textwidth,angle=0]{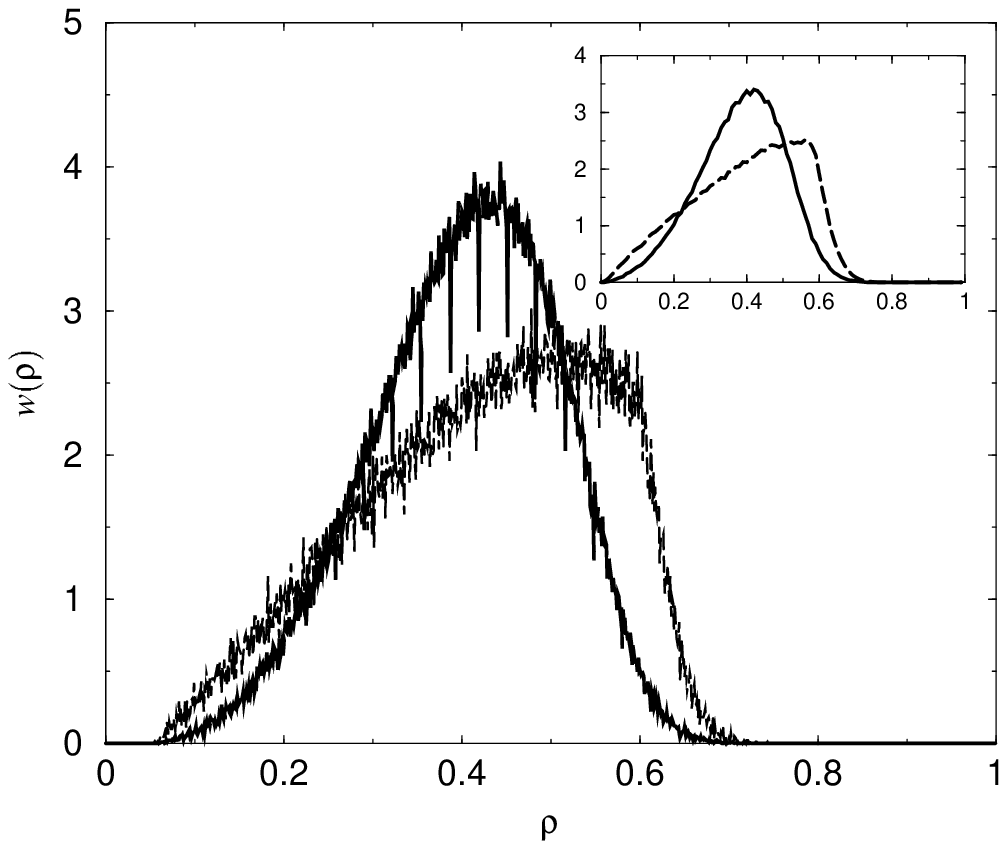}
\caption{$w(\rho)$ for sys\-tems with the parameters
${s_0}/{L^2}$=0.05 and $L=$ $63$, $1600$ (as the peaks move from left
to right). The measured mean tree densities are $\bar{\rho}=$ $0.402$ and
$0.423$. The inset shows $w(\rho)$ for the test simulation with an
exponential probability distribution for $s_0$. The values of $L$, and the mean value of $s_0$ are the same as in the main figure, the measured mean tree densities are 
$\bar{\rho}=$ $0.39$ and $0.408$. 
}
\label{bild8}
\end{figure}

\begin{figure}
\includegraphics[width=0.46\textwidth,angle=0]{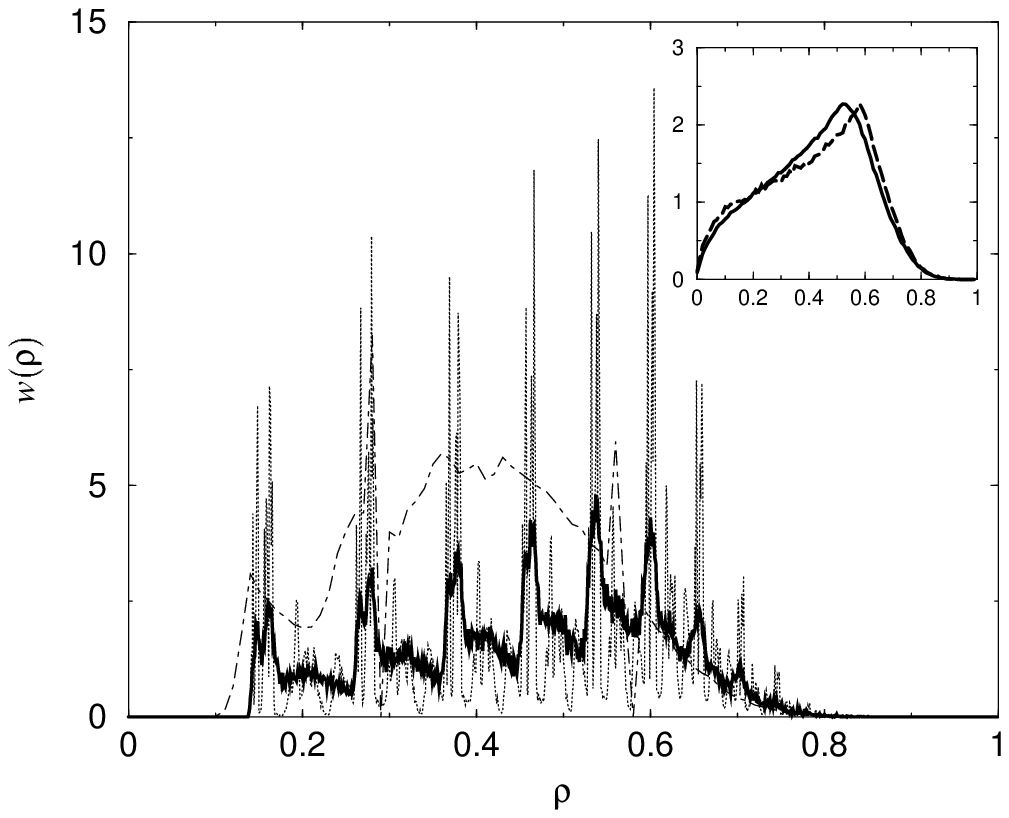}
\caption{$w(\rho)$ for sys\-tems with
the parameters ${s_0}/{L^2}$=0.15 and $L=115$ (for the solid line), 
$L=1600$ (for the dotted line),  $L=10$ (for the dash-dot\-ted line)
The $w(\bar{\rho})$ values for the $L=10$ sys\-tem are mul\-ti\-pli\-ed by 2.
The measured mean tree densities 
are $\bar{\rho}=$
$0.454$, $0.455, 0.394$.  The inset shows $w(\rho)$ for the test simulation with an
exponential probability distribution for $s_0$. The values of $L$, and the mean value of $s_0$ are the same as for the solid and dotted curve in the main figure, the measured mean tree densities are 
$\bar{\rho}=$ $0.425$ and $0.428$. 
}
\label{bild9}
\end{figure}

\begin{figure}
\includegraphics[width=0.46\textwidth,angle=0]{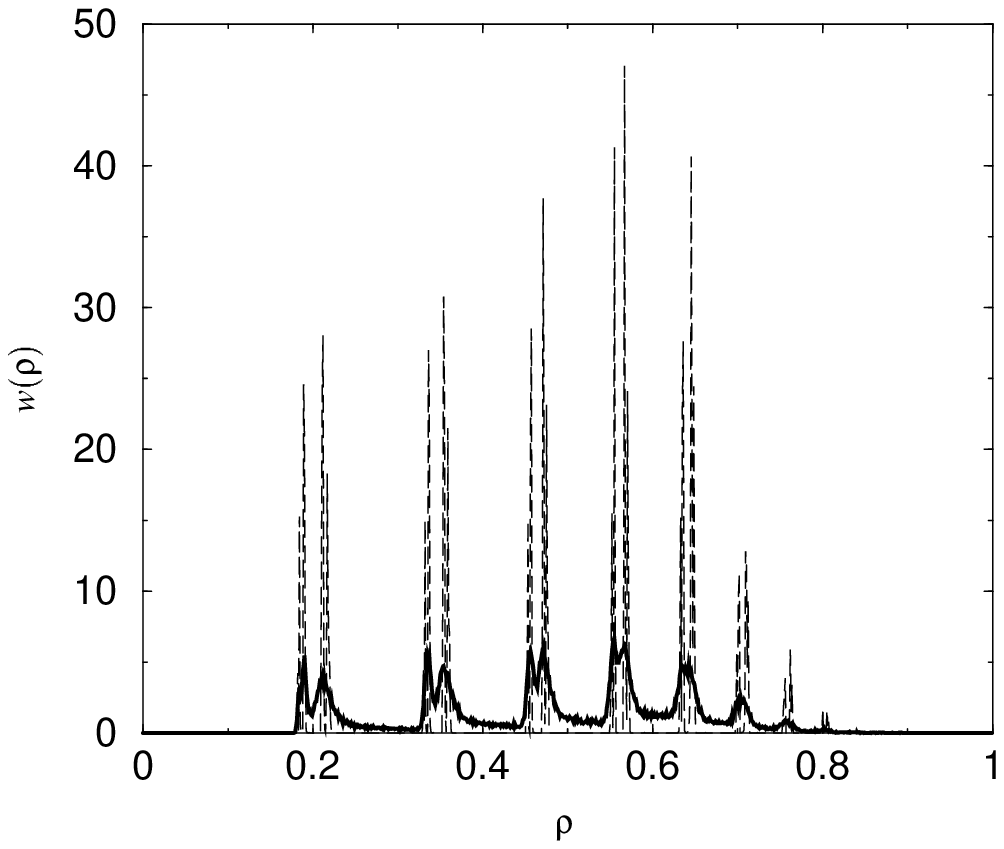}
\caption{$w(\rho)$ for sys\-tems with the parameters
${s_0}/{L^2}$=0.20 and $L=$ $100$ (for the solid line), $1600$ (for the
dot\-ted line).  The measured mean tree densities  are $\bar{\rho}=$ 
$0.474$, $0.473$. 
} \label{bild10}
\end{figure}

In Figures~\ref{bild7} to~\ref{bild10} $w(\rho)$ is shown for
different values for $L$ and $s_0$. For large enough and fixed
$s_0/L^2$, the mean tree density increases with increasing system
size, until it reaches its asymptotic value above $\bar\rho_c$. For
fixed $L$, the mean tree density increases with increasing $s_0$.
Apart from this increase in mean tree density, the following other
trends are observed: (a) As the mean density approaches $\bar \rho_c$,
the curves for $w(\rho)$ become broader (Figure~\ref{bild7}). This is
because the patches of homogeneous density visible in
Figure~\ref{bild1} become larger with increasing $\bar \rho$, leading
to larger global density fluctuations.  (b) As the mean tree density
increases above $\bar \rho_c$, the shape of the distribution becomes
asymmetric, with the maximum moving from $\bar \rho_c$ to the
percolation threshold $\rho_{perc} \simeq 0.59$
(Figure~\ref{bild8}). The reason is that for $\bar \rho > \bar \rho_c$
the patchy structure is replaced by a more homogeneous (percolation
like) structure, where the largest fires are system spanning and occur
for densities above the percolation threshold. Once the density lies
above the percolation threshold the probability that a system spanning
fire occurs is very high.  This is why densities much higher than the
percolation occur seldom, explaining the rapid decrease of $w(\rho)$
above the percolation threshold.  (c) As the system size increases for
fixed $p=s_0/L^2$, there occur peaks in the density distribution which
become sharper and more numerous for larger $L$ (Figures~\ref{bild9}
and \ref{bild10}). This can be explained by realizing that the
difference between finite and system spanning fires becomes more
pronounced as $L$ increases. In the limit $L\to \infty$, finite fires
do not affect the density at all, while system spanning fires reduce
it to a small value. Subsequent filling events then increase the
density to $1-\exp(-p)$, $1-\exp(-2p)$, $1-\exp(-3p)$, etc., until the
density is above the percolation threshold and another system spanning
fire can occur. These system spanning fires do not always occur at the
first instance where the density is above $\rho_{perc}$, since
lightning might strike and empty a site. Also, the density immediately
after a system spanning fire depends slightly on the density before
the fire. Therefore, the series of density values given above, becomes
slightly shifted, depending on the density just before the last system
spanning fire. These shifted series of peaks, in turn, give rise to
further possible density values above $\rho_{perc}$, leading to an
additional series of peaks, etc. This is the mechanism leading to the
fractal peak structure that emerges as $L$ is increased. For smaller
$L$, the effect of small fires leads to a larger width of the peaks,
which can therefore not be resolved when they are close together.

As mentioned in the introduction, the peaks in $w(\rho)$ are due to
the fact that lightning can strike the system only between two filling
steps. Had we instead performed our simulations using a small tree
growth probability $p$ and a small lightning probability $f$,
lightning could strike the system between the growth of any two
trees. However, such a simulation would be very slow. In oder to make
sure that our choice of the algorithm has no other effect on the
results apart from the peaks in $w(\rho)$, we performed a test
simulation where $s_0$ is not the same for each filling step. For each
filling step, we chose $s_0$ randomly from an exponential distribution
$P(s_0)=(L^2p)^{-1}\exp(-s_0/(L^2p))$. Such an exponential distribution
results when lightning can strike the system between the growth of any
two trees with the same probability. The mean number of trees growing
between two lightning strokes is now smaller than before. The reason
is that the majority of filling steps increase the tree number by a
value smaller than $L^2p$, and that during large filling events the
tree density becomes high and most of the sites chosen for filling are
already occupied.  The mean tree densities evaluated according to
Eq.~(\ref{rho}) are consequently slightly smaller than before.

The probability density $w(\rho)$ for the tree
density resulting from this modified algorithm is shown in the insets
in Figures ~\ref{bild8} and~\ref{bild9}. As expected, the peaks have
vanished, while the change in shape from a curve with peak around 0.4
to a curve with peak near 0.6 due to finite-size effects is the same
as before.

\section{Conclusion}

In this paper, we have studied finite-size effects in the SOC
forest-fire model.  As these effects become stronger, the system
rearranges from a structure with patches of different densities to a
more homogeneous structure with large density fluctuations in time.
This rearrangement is reflected in the structure of the fire size
distribution, in the mean tree density, and in the temporal density
fluctuations. Qualitatively similar (although quantitatively
different) rearrangements are observed when smaller and smaller
sections of a large SOC system are studied. Due to these qualitative
changes, conventional finite-size scaling does not hold. Our work thus
demonstrates that concepts from equilibrium critical phenomena cannot
be taken over to the study of SOC systems such as the forest-fire
model. Instead, these nonequilibrium critical systems show generically
new features unknown in equilibrium. As the scaling ansatz
eq.~(\ref{scaling}) which is based on a single length scale $\xi \sim
s_0^\nu$ can only be approximately correct, the true asymptotic
scaling behaviour of the model is still an open question.

We suggest that the reason for the unconventional behaviour of the SOC
forest-fire model is the fact that two qualitatively different types
of fires occur: those that burn down a patch of high tree density of
fractal dimension 2, and those that burn down a tree cluster of a
smaller fractal dimension within a region of a tree density below the
percolation threshold. As a consequence, the scaling behaviour of the
system cannot be characterized using only one length scale. While the
superposition of the two types of fires creates the impression of
simple scaling as long as finite-size effects are small, the
difference between them becomes clearly visible for smaller system
sizes, where system spanning fires receive a larger weight. We suggest
that the superposition of the two types of fires is also responsible
for the other unconventional features of the SOC forest-fire model
listed in the introduction.

Models related to the present one have been studied in
 \cite{cla95,cla97a} and \cite{cla97}. In these models, the tree
 density is globally conserved by filling exactly the same number of
 trees into the system that have been burnt. As long as the density is
 below the critical value, these models are equivalent and show the
 critical behaviour of the SOC forest-fire model as the critical
 density is approached from below. They were introduced for the
 purpose of studying the SOC forest-fire model beyond the critical
 point, i.e. for densities larger than the critical density $\bar
 \rho_c$. As the density increases beyond the critical density, both
 models undergo large-scale rearrangements. In \cite{cla95,cla97a},
 where trees are refilled only after the end of a fire, the new
 structure consists of a finite number of large domains of different
 density. In \cite{cla97}, where each tree is refilled into the system
 immediately after it is burnt, the new structure has a continuously
 burning fire, and resembles the forest-fire model without lightning
 introduced earlier by Bak, Chen, and Tang \cite{bak90}, which shows
 spiral-shaped fire fronts \cite{gra91}. In both these models, the
 dynamics in the restructured state are dominated by large fires
 burning forests of a fractal dimension two, similarly to the
 restructuring due to finite-size effects reported in this paper.
 
 Let us conclude by noting that it is unclear whether the behaviour in
 higher dimensions resembles that in two dimensions. Clearly, as long
 as the ``patchy'' structure with two qualitatively different types of
 fire occurs, mean-field theory which neglects all spatial structure
 \cite{chr93,dro94c,ves98} cannot apply, and the system must be below
 its upper critical dimension. The recent paper by Br\"oker and
 Grassberger \cite{bro97} on the forest-fire model without lightning
 indicates that unusual scaling behaviour can occur also in 3 and 4
 dimensional forest-fire models. If 6 is the upper critical dimension
 of the forest-fire model, as suggested in \cite{cla94,chr93,dro94c},
 then the scaling behaviour of the forest-fire model should be
 conventional above 6 dimensions.

\acknowledgements
This work was supported by EPSRC Grant
GR/K79307, and by the EU network project (TMR) ``fractal
structures and self organization'', EU-contract ERPFMRXCT980183.

\newpage

\vskip 0.7\textwidth

\end{multicols}

\end{document}